\documentclass[12pt]{article}
\usepackage{latexsym,amsfonts,amssymb}
\makeatletter
\@addtoreset{equation}{subsection}
\makeatother

\begin{document}
\begin{center}
{\Large \bf The Eternal Closed Universe:\\Deflation-Inflation}
\\[1.5cm]
 {\bf Vladimir S.~MASHKEVICH}\footnote {E-mail:
  Vladimir.Mashkevich100@qc.cuny.edu}
\\[1.4cm] {\it Physics Department
 \\ Queens College\\ The City University of New York\\
 65-30 Kissena Boulevard\\ Flushing, New York
 11367-1519} \\[1.4cm] \vskip 1cm

{\large \bf Abstract}
\end{center}

An outgrowth of the idea of inflation is advanced. In the inflation regime, the singularity condition is broken. Equations which govern inflation are invariant under time reversal, so that they describe deflation as well. Those two observations suggest that inflation may be extended to the following process: deflation ($t<0$)---minimum radius $>0$ ($t=0$)---inflation ($t>0$), with no singularity. A relevant construction is carried out in the framework both of classical and of reductive semiclassical gravity. The construction results in an eternal ($-\infty<t<\infty$) contracting-expanding closed universe.

\newpage

\section*{Introduction}
In modern cosmology, it is generally taken that an essential stage
in the time evolution of the universe is inflation [1-4].
Inflation is defined by the inequality
$\ddot{R}:=d^{2}R/dt^{2}>0$, which, in view of the so-called Raychaudhuri equation [5] $\ddot{R}/R=-(4\pi\varkappa/3)(\varrho+3p)$, reduces to $\varrho+3p<0$. ($R$ is the scale factor, or the radius of the universe, $\varrho$ is the energy density, $p$ is the pressure, and $\varkappa=t^{2}_{\mathrm{Planck}}$ is the gravitational constant, $c=\hbar=1.$)

In the simplest model ($-p=\varrho=\mathrm{const}>0$) [2a, 3],
$R(t)=R_{\mathrm{end}}\exp\{H(t-t_{\mathrm{end}})\},\;t<t_{\mathrm{end}
}\,, H=\sqrt{(8\pi\varkappa/3)\varrho}$\,, where  the
subscript $\mathrm{end}$ stands for the end of inflation. This
result is obtained as a solution to the Friedmann equation,
$\dot{R}^{2}+k=H^{2}R^{2}$, with $k=0$ (flat universe) or
$k=1$ (closed universe) and $H^{2}R^{2}\gg k$.

The idea of inflation has been introduced to overcome three classic cosmological problems [1]: flatness, horizons, monopoles. However, one of the most poignant problems that challenge cosmology is that of singularity [2].

For the FLRW universe, the singularity condition is of the form $\varrho+3p>0$ [5-7].
In the inflation regime, this condition is broken.

Next, equations which govern inflation are invariant under time reversal, so that they describe deflation as well.

Those two observations suggest that inflation may be extended to the following process: deflation ($t<0$)---$R_{\mathrm{minimum}}(0)>0$---inflation ($t>0$), where there is no singularity.

Indeed, for the closed universe, there exists an exact solution to the Friedmann equation, $R(t)=(1/H)\cosh Ht$, which describes the above process. This solution, to be sure, has been known [2a], which, however, has not brought about the idea of deflation: the solution has been considered only for $Ht\gg 1$ when $R(t)\propto \mathrm{e}^{Ht}$ (inflation).

In the present paper, a construction based on the above extension is carried out in the framework both of classical and of reductive semiclassical gravity [8] (the latter takes into account pseudomatter, which represents dark matter). Deflation-inflation is due to phase transitions in superdense matter, after which symmetry between the strong and electroweak interactions is restored/broken and a scalar field is created/annihilated. The energy density and pressure of the field $\varphi$ with potential $V(\varphi)$ are: $\varrho=(1/2)\dot{\varphi}^{2}+V(\varphi),\;\;p=(1/2)\dot{\varphi}^{2}-V(\varphi)$, so that $\varrho+3p=2[\dot{\varphi}^{2}-V(\varphi)]<0$. In the deflation-inflation regime, the symmetry holds and the inequality $\dot{\varphi}^{2}-V(\varphi)<0$ is fulfilled.

The construction results in an eternal ($-\infty<t<\infty$) contracting-expanding closed universe with no singularity at $t=0$ (or any other time). The closure of the universe is essential; it is the consequence of the continuity of $\dot{R}$ at $t=0$.

Sequential stages of time evolution are these:
\begin{eqnarray}
&&\Lambda:\,(-\infty,-t_{-}),\dot{R}<0,\ddot{R}>0;\quad\mathrm{perfect\;fluid}:\,
(-t_{-},-t_{+}),\dot{R}<0,\ddot{R}<0;\nonumber\\
&&\mathrm{scalar\;field}:\,(-t_{+},t_{+}),-\dot{R}(t_{+})<\dot{R}<\dot{R}(t_{+}),
\ddot{R}>0;\nonumber\\
&&\mathrm{perfect\;fluid}:\,
(t_{+},t_{-}),\dot{R}>0,\ddot{R}<0;\quad\Lambda :\,(t_{-},\infty),\dot{R}>0,\ddot{R}>0\nonumber
\end{eqnarray}
where $\Lambda$ is the cosmological constant.

\section{The Robertson-Walker metric and matter}

\subsection{The RW metric}

The standard models of modern cosmology are based on the assumption that the universe is isotropic about every point, which implies homogeneity. The resulting metric is the RW one:
\begin{equation}%(1.1.1)
\mathrm{d}s^{2}=\mathrm{d}t^{2}-R^{2}(t)\left[\frac{\mathrm{d}r^{2}}{1-kr^{2}}
+r^{2}(\mathrm{d\theta}^{2}+\sin^{2}\theta\,\mathrm{d}\phi^{2})\right],\qquad k=1,0,-1
\end{equation}

\subsection{The energy-momentum tensor and matter equation}

The symmetry of spacetime results in the energy-momentum tensor of the form [1]
\begin{equation} % (1.2.1)
T_{\mu\nu}=(\varrho+p)u_{\mu}u_{\nu}-pg_{\mu\nu}\,,\quad \mu,\nu=0,1,2,3
\end{equation}
with
\begin{equation} % (1.2.2)
u_{i}=0,\quad i=1,2,3,\qquad u_{0}^{2}=u_{0}u^{0}=1
\end{equation}

The matter equation is of the form
\begin{equation} % (1.2.3)
\frac{\mathrm{d}}{\mathrm{d}t}(\varrho R^{3})=-p\frac{\mathrm{d}R^{3}}{\mathrm{d}t}
\end{equation}
or
\begin{equation} % (1.2.4)
R\dot{\varrho}+3(\varrho+p)\dot{R}=0
\end{equation}

\section{Deflation-inflation}

\subsection{Definition}

The deflation-inflation stage of dynamics is defined by the following conditions:
\begin{equation} % (2.1.1)
f(-t)=f(t),\quad f=R,\varrho,p
\end{equation}
\begin{equation} % (2.1.2)
\ddot{R}(t)>0\quad \mathrm{for}\;\;t\in(-t_{+},t_{+}),\;\;t_{+}>0
\end{equation}
\begin{equation} % 2.1.3
R(0)=:R_{0}=R_{\mathrm{minimum}}>0
\end{equation}

\subsection{Conditions for time derivatives at $t=0$}

From (2.1.1) follows
\begin{equation} % 2.2.1
\dot{f}(0)=0,\quad f=R,\varrho,p
\end{equation}

\subsection{Relations between $p$ and $\varrho$ and dynamical equations for matter}

According to grand unified theories, in superdense matter phase transitions occur, after which symmetry between the strong and electroweak interactions is restored/broken, with the result that a scalar field $\varphi$ is created/annihilated [2]. The energy density and pressure of a spatially homogeneous scalar field $\varphi(t)$ with potential $V(\varphi)$ in the RW spacetime take the form [1]
\begin{equation} % 2.3.1
\varrho=\frac{1}{2}\dot{\varphi}^{2}+V(\varphi),\;\;p=\frac{1}{2}\dot{\varphi}^{2}-V(\varphi)
\end{equation}
In the deflation-inflation regime, the symmetry holds, so that pressure and density are related via the scalar field by (2.3.1).

The matter equation (1.2.4) reduces to
\begin{equation} % 2.3.2
\dot{\varphi}[R\ddot{\varphi}+3\dot{R}\dot{\varphi}+RV'(\varphi)]=0
\end{equation}
where $V'(\varphi)=\mathrm{d}V/\mathrm{d}\varphi$. Thus, the dynamical equation for matter is either
\begin{equation} % 2.3.3
\dot{\varphi}=0
\end{equation}
or
\begin{equation} % 2.3.4
R\ddot{\varphi}+3\dot{R}\dot{\varphi}+RV'(\varphi)=0
\end{equation}
If both (2.3.4) and (2.3.3) hold, then
\begin{equation} % 2.3.5
V'(\varphi)=0
\end{equation}
From (2.3.3), (2.3.1) follows
\begin{equation} % 2.3.6
p=-\varrho=\mathrm{const}
\end{equation}

Conditions (2.1.1), (2.2.1) reduce to
\begin{equation} % 2.3.7
R(-t)=R(t),\quad\varphi(-t)=\varphi(t)
\end{equation}
\begin{equation} % 2.3.8
\dot{R}(0)=0,\quad \dot{\varphi}(0)=0
\end{equation}

In the regime outside of deflation-inflation, $p$ and $\varrho$ are related by a state equation,
\begin{equation} % 2.3.9
p=p(\varrho)
\end{equation}
The dynamical equation for matter is (1.2.4) or the second order equation:
\begin{equation} % 2.3.10
R\ddot{\varrho}+3(\varrho+p)\ddot{R}+\left(1+3\frac{\mathrm{d}p}{\mathrm{d}\varrho}\right)
\dot{R}\dot{\varrho}=0
\end{equation}

\newpage

\section{Dynamics in classical gravity}

\subsection{Dynamical equations and constraint}

The Einstein equations for the FLRW universe are [9]:
\begin{equation} % 3.1.1
2R\ddot{R}+\dot{R}^{2}+k-\Lambda R^{2}+8\pi\varkappa pR^{2}=0
\end{equation}
\begin{equation} % 3.1.2
3(\dot{R}^{2}+k)-\Lambda R^{2}-8\pi\varkappa \varrho R^{2}=0
\end{equation}
where $\Lambda$ is the cosmological constant. The dynamical equation is (3.1.1), and (3.1.2) is a constraint on initial conditions [10]. Another dynamical equation is the one for matter (Subsection 1.2).

Note that had (3.1.2), (1.2.3) been used as dynamical equations [in place of (3.1.1), (1.2.3)], for the solution
\begin{equation} % 3.1.3
R=\mathrm{const},\quad \varrho=\frac{1}{8\pi\varkappa}\left(\frac{3k}{R^{2}}-\Lambda\right)
\end{equation}
the result
\begin{equation} % 3.1.4
p=-\frac{1}{8\pi\varkappa}\left(\frac{k}{R^{2}}-\Lambda\right)
\end{equation}
would have failed. The point is that (3.1.1) follows from (3.1.2), (1.2.3) if and only if $\dot{R}=0$.

\subsection{Deflation-inflation stage. Closed universe}

For the deflation-inflation stage of time evolution, from (3.1.1), (3.1.2), and (2.3.1) follows
\begin{equation} % 3.2.1
\ddot{R}=\Gamma^{2}R
\end{equation}
where
\begin{equation} % 3.2.2
\Gamma^{2}=-\frac{8\pi\varkappa}{3}[\dot{\varphi}^{2}-V(\varphi)]+\frac{1}{3}\Lambda
\end{equation}
so that
\begin{equation} % 3.2.3
\Gamma^{2}>0\quad\mathrm{for}\;t\in(-t_{+},t_{+})
\end{equation}
must hold.

From (3.1.2) at $t=0$ follows
\begin{equation} % 3.2.4
R_{0}^{2}=\frac{3k}{\Lambda+8\pi\varkappa\varrho_{0}}=
\frac{3k}{\Lambda+8\pi\varkappa V(\varphi_{0})}=\frac{k}{\Gamma_{0}^{2}}
\end{equation}
Thus,
\begin{equation} % 3.2.5
k=1\qquad(\mathrm{closed\;universe})
\end{equation}
---deflation-inflation is possible only in a closed universe. The closure condition is the consequence of the continuity of $\dot{R}$ at $t=0$. For $k=0$, we would obtain from (3.1.2) $-\dot{R}(-0)=\dot{R}(+0)>0$, which would result in $\ddot{R}(0)\propto \delta(0)$ in (3.1.1).

\newpage

The dynamical equations are:
\begin{equation} % 3.2.6
2R\ddot{R}+\dot{R}^{2}+1-\Lambda R^{2}+8\pi\varkappa\left[\frac{1}{2}
\dot{\varphi}^{2}-V(\varphi)\right]R^{2}=0
\end{equation}
\begin{equation} % 3.2.7
R\ddot{\varphi}+3\dot{R}\dot{\varphi}+RV'(\varphi)=0\qquad\mathrm{or}\quad
\dot{\varphi}=0
\end{equation}
with the initial conditions:
\begin{equation} % 3.2.8
\varphi(0)=\varphi_{0},\;\;R_{0}=\frac{1}{\Gamma_{0}},\qquad\dot{\varphi}(0)=0,\;\;
\dot{R}(0)=0
\end{equation}

The constraint (3.1.2) is satisfied at $t=0$.

\subsection{Dynamics outside of deflation-inflation stage}

Dynamical equations for
\begin{equation} % 3.3.1
t\in(-\infty,-t_{+}]\cup[t_{+},+\infty)
\end{equation}
are (3.1.1) and (1.2.4) [or (2.3.10)] with the constraint (3.1.2) and the state equation (2.3.9). Initial conditions at $t=t_{+}$ are:
\begin{equation} % 3.3.2
R(t_{+})=R(t_{+}-0),\;\;\dot{R}(t_{+})=\dot{R}(t_{+}-0),\quad
\varrho(t_{+})=\varrho(t_{+}-0)=\frac{1}{2}[\dot{\varphi}(t_{+}-0)]^{2}+
V(\varphi(t_{+}-0))
\end{equation}
[If (2.3.10) is used, $\dot{\varrho}(t_{+})$ is determined by the constraint (1.2.4).] The constraint (3.1.2) is satisfied at $t=t_{+}$.

\subsection{The simplest model: $\Gamma$=const}

Put
\begin{equation} % 3.4.1
\dot{\varphi}^{2}-V(\varphi)=-b^{2}=\mathrm{const}\quad\mathrm{for}\;\;
t\in(-t_{+}',t_{+}'),\;\;0<t_{+}'\leq t_{+}
\end{equation}
whence follows
\begin{equation} % 3.4.2
[2\ddot{\varphi}-V'(\varphi)]\dot{\varphi}=0
\end{equation}
The equation
\begin{equation} % 3.4.3
2\ddot{\varphi}-V'(\varphi)=0
\end{equation}
would result, in view of (3.2.7), in two equations for $\varphi$, so that we have
\begin{equation} % 3.4.4
\dot{\varphi}=0,\quad\varphi=\varphi_{0}\quad b^{2}=V(\varphi_{0})
\end{equation}
and
\begin{equation} % 3.4.5
V'(\varphi_{0})=0
\end{equation}

Now
\begin{equation} % 3.4.6
\Gamma^{2}=\frac{8\pi\varkappa}{3}V(\varphi_{0})+\frac{1}{3}\Lambda
\end{equation}
and we obtain
\begin{equation} % 3.4.7
R=\frac{1}{\Gamma}\cosh\Gamma t,\quad\dot{R}=\sinh\Gamma t\qquad\mathrm{for}\;
t\in(-t_{+}',t_{+}')
\end{equation}

\subsection{The $\Lambda$ stage}

Since
\begin{equation}\label{3.5.1}
\varrho R^{2}\rightarrow 0,\quad pR^{2}\rightarrow 0 \qquad
\mathrm{for}\;R\rightarrow \infty
\end{equation}
equations (3.1.1), (3.1.2) for $R\rightarrow \infty$ reduce to
\begin{equation}\label{3.5.2}
2R\ddot{R}+\dot{R}^{2}+1-\Lambda R^{2}=0
\end{equation}
\begin{equation}\label{3.5.3}
3(\dot{R}^{2}+1)-\Lambda R^{2}=0
\end{equation}
whence follows
\begin{equation}\label{3.5.4}
\ddot{R}-\frac{1}{3}\Lambda R=0
\end{equation}
so that
\begin{equation}\label{3.5.5}
R=c_{+}\mathrm{e}^{\sqrt{\Lambda/3}\;t}+c_{-}\mathrm{e}^{-\sqrt{\Lambda/3}\;t}
\end{equation}
and by the constraint (3.5.3)
\begin{equation}\label{3.5.6}
R=\frac{1}{\sqrt{\Lambda/3}}\cosh\sqrt{\Lambda/3}\;t\,,\quad
\dot{R}=\sinh\sqrt{\Lambda/3}\;t\qquad\mathrm{for}\;t\rightarrow\mp\infty
\end{equation}

\subsection {The conventional inflation}

In the framework of the simplest model, the conventional inflation
corresponds to
\begin{equation}\label{3.6.1}
\Lambda=0,\quad
H=\Gamma=\sqrt{\frac{8\pi\varkappa\varrho_{0}}{3}}\,,\quad \Gamma
t\gg 1
\end{equation}
\begin{equation}\label{3.6.2}
R(t)=\frac{1}{H}\mathrm{e}^{Ht}
\end{equation}
Thus, we have constructed an extension of the conventional
inflation.

\section{Dynamics in reductive semiclassical gravity}

\subsection{Dynamical equations, constraint, and pseudomatter}

The extended Einstein equations for the FLRW universe are [8]:
\begin{equation}%4.1.1
2R\ddot{R}+\dot{R}^{2}+k-\Lambda R^{2}+8\pi\varkappa pR^{2}=0
\end{equation}
\begin{equation}%4.1.2
3(\dot{R}^{2}+k)-\Lambda R^{2}-8\pi\varkappa (\varrho+\varepsilon)
R^{2}=0
\end{equation}
where $\varepsilon$ is the density of pseudomatter, which
represents dark matter.

From (4.1.1), (4.1.2) follows
\begin{equation}%4.1.3
\frac{\mathrm{d}}{\mathrm{d}t}[(\varrho+\varepsilon)
R^{3}]=-p\frac{\mathrm{d}R^{3}}{\mathrm{d}t}
\end{equation}
which, in view of (1.2.3), results in
\begin{equation} % 4.1.4
\varepsilon=\frac{B}{R^{3}},\quad B=\mathrm{const}
\end{equation}

In view of (4.1.3) and (1.2.3), the only possibility for $\varepsilon$ is (4.1.4), so that (4.1.3) is, in fact, independent of the Einstein equations. Therefore the dynamical equation is (4.1.1), whereas (4.1.2) is a constraint on initial conditions. Another dynamical equation is that for matter (Subsection 2.3).

\subsection{Deflation-inflation stage. Closed universe}

From (4.1.1), (4.1.2), and (2.3.1) follows
\begin{equation} % 4.2.1
\ddot{R}=\Gamma^{2}R
\end{equation}
with
\begin{equation} % 4.2.2
\Gamma^{2}=-\frac{4\pi\varkappa}{3}[(\rho+\varepsilon)+3p]+\frac{1}{3}\Lambda=-\frac{4\pi\varkappa}{3}
\left\{2[\dot{\varphi}^{2}-V(\varphi)]+\frac{B}{R^{3}}\right\}+\frac{1}{3}\Lambda
\end{equation}
The inequality
\begin{equation} % 4.2.3
\Gamma^{2}>0\qquad\mathrm{for}\;\;t\in(-t_{+},t_{+})
\end{equation}
must hold.

From (4.1.2) at $t=0$ follows
\begin{equation} % 4.2.4
R_{0}^{2}=\frac{3k}{\Lambda+8\pi\varkappa(\varrho_{0}+\varepsilon_{0})}
=\frac{3k}{\Lambda+8\pi\varkappa[V(\varphi_{0})+B/R_{0}^{3}]}
\end{equation}
Thus, $k\neq 0$ and, since $k=-1$ is of no interest,
\begin{equation} % 4.2.5
k=1\qquad(\mathrm{closed\;universe})
\end{equation}
i.e., deflation-inflation is possible only in a closed universe.

In fact, (4.2.4) determines $B$:
\begin{equation} % 4.2.6
B=R_{0}\left[\frac{3-R_{0}^{2}\Lambda}{8\pi\varkappa}-
R_{0}^{2}V(\varphi_{0})\right]
\end{equation}

The dynamical equations are:
\begin{equation} % 4.2.7
2R\ddot{R}+\dot{R}^{2}+1-\Lambda R^{2}+
8\pi\varkappa\left[\frac{1}{2}\dot{\varphi}^{2}-V(\varphi)\right]R^{2}=0
\end{equation}
\begin{equation} % 4.2.8
R\ddot{\varphi}+3\dot{R}\dot{\varphi}+RV'(\varphi)=0\qquad\mathrm{or}\;\;
\dot{\varphi}=0
\end{equation}
with the initial conditions:
\begin{equation} % 4.2.9
\varphi(0)=\varphi_{0},\;\;\dot{\varphi}(0)=0,\qquad
R(0)=R_{0},\;\;\dot{R}(0)=0
\end{equation}
The constraint (4.1.2) is satisfied at $t=0$.

The pseudomatter density is:
\begin{equation} % 4.2.10
\varepsilon=\frac{B}{R^{3}}
\end{equation}
where $B$ is given by (4.2.6).
\subsection{Dynamics outside of deflation-inflation stage}

Dynamical equations for
\begin{equation} % 4.3.1
t\in(-\infty,-t_{+}]\cup[t_{+},+\infty)
\end{equation}
are (4.1.1) and (1.2.4) [or (2.3.10)] with the constraint (4.1.2) and the state equation (2.3.9). Initial conditions for $t=\pm t_{+}$ are:
\begin{eqnarray}
R(\pm t_{+})=R(\pm t_{+}\mp 0),\;\;\dot{R}(\pm t_{+})=\dot{R}(\pm t_{+}\mp 0),\nonumber\\
\varrho(\pm t_{+})=\varrho(\pm t_{+}\mp 0)=\frac{1}{2}[\dot{\varphi}(\pm t_{+}\mp 0)]^{2}+
V(\varphi(\pm t_{+}\mp 0))
\end{eqnarray}
The constraint (4.1.2) is satisfied at $t=\pm t_{+}$.

The pseudomatter density is:
\begin{equation} % 4.3.3
\varepsilon=\frac{B}{R^{3}}
\end{equation}
with $B$ given by (4.2.6).

\subsection{On the model with $\Gamma$=const}

Consider the possibility of the equality
\begin{equation} % 4.4.1
\varrho+\varepsilon+3p=-b=\mathrm{const}
\end{equation}
in which case
\begin{equation} % 4.4.2
\Gamma^{2}=\mathrm{const}
\end{equation}
From (4.1.3) we obtain
\begin{equation} % 4.4.3
\dot{p}R+(b+2p)\dot{R}=0
\end{equation}
or
\begin{equation} % 4.4.4
\frac{\mathrm{d}p}{b+2p}+\frac{\mathrm{d}R}{R}=0
\end{equation}
whence
\begin{equation} % 4.4.5
p=\frac{a}{R^{2}}-\frac{b}{2}
\end{equation}
Now, from (4.4.1) and (4.4.5) follows
\begin{equation} % 4.4.6
(\varrho+p)+\frac{B}{R^{3}}+\frac{2a}{R^{2}}=0
\end{equation}
i.e.,
\begin{equation} % 4.4.7
\dot{\varphi}^{2}+\frac{B}{R^{3}}+\frac{2a}{R^{2}}=0
\end{equation}
We have (4.2.8) and (4.4.7): two equations for $\varphi$ [if $\Gamma$=const, then $R=(1/\Gamma)\cosh\Gamma t$]. Thus, (4.4.1) is impossible.

Now consider the case
\begin{equation} % 4.4.8
\dot{\varphi}=0,\quad\varphi=\varphi_{0},\quad V'(\varphi_{0})=0\qquad\mathrm{for}\;\,
t\in(-t'_{+},t'_{+}),\;\,0<t'_{+}\leq t_{+}
\end{equation}
The dynamical equation (4.1.1) reduces to
\begin{equation} % 4.4.9
2R\ddot{R}+\dot{R}^{2}+1-\Lambda R^{2}-8\pi\varkappa V(\varphi_{0})R^{2}=0
\end{equation}
Equivalently, from (4.1.1), (4.1.2) we obtain
\begin{equation} % 4.4.10
\ddot{R}=\left[\frac{8\pi\varkappa}{3}V(\varphi_{0})+\frac{1}{3}\Lambda\right]R-
\frac{4\pi\varkappa B}{3R^{2}}
\end{equation}
with
\begin{equation} % 4.4.11
R(0)=R_{0},\quad\dot{R}(0)=0
\end{equation}

If
\begin{equation} % 4.4.12
B\ll 2V(\varphi_{0})R_{0}^{3}
\end{equation}
then
\begin{equation} % 4.4.13
R(t)\approx \bar{R}(t)
\end{equation}
\begin{equation} % 4.4.14
\bar{R}(t)=\frac{1}{\bar{\Gamma}}\coth\bar{\Gamma}t,\quad
\bar{\Gamma}^{2}=\frac{8\pi\varkappa}{3}V(\varphi_{0})+\frac{1}{3}\Lambda
\end{equation}

Note that (4.4.9) is satisfied by (4.4.14), but (4.4.2) is satisfied only if $B=0$.

\subsection{The $\Lambda$ stage}

We have
\begin{equation}\label{4.5.1}
\varrho R^{2}\rightarrow 0,\quad \frac{B}{R^{3}}R^{2}\rightarrow 0
,\quad pR^{2}\rightarrow 0\qquad\mathrm{for} R\rightarrow\infty
\end{equation}
so that (4.1.1), (4.1.2) for $R\rightarrow\infty$ reduce to
\begin{equation}\label{4.5.2}
2R\ddot{R}+\dot{R}^{2}+1-\Lambda R^{2}=0
\end{equation}
\begin{equation}\label{4.5.3}
3(\dot{R}^{2}+1)-\Lambda R^{2}=0
\end{equation}
whence it follows that
\begin{equation}\label{4.5.4}
R=\frac{1}{\sqrt{\Lambda/3}}\cosh\sqrt{\Lambda/3}\;t\,,\quad
\dot{R}=\sinh\sqrt{\Lambda/3}\;t\qquad\mathrm{for}\;t\rightarrow\mp\infty
\end{equation}

\section{Dynamics of the eternal closed universe}

\subsection{Stages of time evolution}

The closed universe with deflation-inflation is eternal with $R(t)>0$. Sequential stages of its time evolution are these:
\begin{eqnarray}
&&\Lambda:\,(-\infty,-t_{-}),\dot{R}<0,\ddot{R}>0;\quad\mathrm{perfect\;fluid}:\,
(-t_{-},-t_{+}),\dot{R}<0,\ddot{R}<0;\nonumber\\
&&\mathrm{scalar\;field}:\,(-t_{+},t_{+}),-\dot{R}(t_{+})<\dot{R}<\dot{R}(t_{+}),
\ddot{R}>0;\nonumber\\
&&\mathrm{perfect\;fluid}:\,
(t_{+},t_{-}),\dot{R}>0,\ddot{R}<0;\quad\Lambda:\,(t_{-},\infty),\dot{R}>0,\ddot{R}>0
\end{eqnarray}

\subsection{No singularity and no quantum gravity epoch}

Since
\begin{equation} % 5.2.1
R_{\mathrm{minimum}}=R(0)>0\quad \mathrm{and}\quad R(t)<\infty \;\;
\mathrm{for}\;-\infty<t<\infty
\end{equation}
there is no singularity at $t=0$ or any other time.

According to the present epoch data,
\begin{equation} % 5.2.2
\varepsilon(t_{+})\ll\varrho(t_{+})
\end{equation}
so that, by (4.4.14),
\begin{equation} % 5.2.3
R_{\mathrm{minimum}}\approx\frac{1}{\bar{\Gamma}}\approx
\sqrt{\frac{3}{8\pi\varkappa V(\varphi_{0})}}=
\sqrt{\frac{3}{8\pi\varkappa\varrho_{0}}}
\end{equation}
Put [1]
\begin{equation} % 5.2.4
\varrho_{0}=(10^{16}\mathrm{GeV})^{4}
\end{equation}
then
\begin{equation} % 5.2.5
R_{\mathrm{minimum}}=2.78\times 10^{-38}\,\mathrm{sec}\gg 5.39\times
10^{-44}\,\mathrm{sec}=t_{\mathrm{Planck}}
\end{equation}
In this sense, there has been no quantum gravity epoch; the entire spacetime is described classically.

\subsection{Initial conditions at $t\rightarrow-\infty$}

We have exploited initial conditions at $t=0$. It may be of interest to relate them to $-\infty$, when $R\rightarrow\infty$.

In reductive semiclassical gravity, the dynamical equations are:
\begin{equation} % 5.3.1
2R\ddot{R}+\dot{R}^{2}+k-\Lambda R^{2}+8\pi\varkappa pR^{2}=0
\end{equation}
\begin{equation} % 5.3.2
\frac{\mathrm{d}}{\mathrm{d}t}(\varrho R^{3})=-p\frac{\mathrm{d}R^{3}}{\mathrm{d}t}
\end{equation}
and the constraint is:
\begin{equation}%5.3.3
3(\dot{R}^{2}+k)-\Lambda R^{2}-8\pi\varkappa (\varrho+\frac{B}{R^{3}})
R^{2}=0
\end{equation}
(In classical gravity, $B=0$.)

From (5.3.2) follows
\begin{equation} % 5.3.4
p=-\frac{1}{3R^{2}}\frac{\mathrm{d}(\varrho R^{3})}{\mathrm{d}R}
\end{equation}
and we need to determine $\varrho(R)$. [$\varrho(R)$ and $p(R)$ imply a state equation, $p=p(\varrho)$.]

We have for a particle energy [7]:
\begin{equation} % 5.3.5
\omega=\sqrt{m^{2}+\left(\frac{\xi}{R}\right)^{2}}
\end{equation}
whence
\begin{equation} % 5.3.6
\omega=m+\frac{1}{2m}\left(\frac{\xi}{R}\right)^{2}\quad (m\neq 0),
\qquad \omega=\frac{\xi}{R}\quad (m=0)\qquad \mathrm{for}\;R\rightarrow\infty\quad
(\mathrm{Big\;Cold})
\end{equation}
We obtain
\begin{equation} % 5.3.7
\varrho=\varrho_{\mathrm{mass}}+\varrho_{\mathrm{massless}}=
\left(\frac{\alpha^{2}}{R^{3}}+\frac{\gamma^{2}}{R^{5}}\right)+
\frac{\beta^{2}}{R^{4}}\quad\mathrm{for}\;R\rightarrow\infty\quad
(\mathrm{Big\; Cold\; Void})
\end{equation}
and
\begin{equation} % 5.3.8
p=\frac{2\gamma^{2}}{3R^{5}}+\frac{\beta^{2}}{3R^{4}}\quad\mathrm{for}\;R\rightarrow\infty
\end{equation}

Now, introduce
\begin{equation} % 5.3.9
t_{\mathrm{in}}:=t_{\mathrm{initial}}<0,\quad|t_{\mathrm{in}}|\rightarrow\infty,\quad
R_{\mathrm{in}}:=R(t_{\mathrm{in}})\rightarrow\infty
\end{equation}
From (5.3.3) follows
\begin{equation} % 5.3.10
\dot{R}_{\mathrm{in}}=-\left\{\frac{1}{3}\left[\Lambda R_{\mathrm{in}}^{2}
+8\pi\varkappa\left(\varrho_{\mathrm{in}}+\frac{B}{R_{\mathrm{in}}^{3}}\right)\right]-1
\right\}^{1/2}
\end{equation}
Denote by $R(t,t_{\mathrm{in}})$ the solution to (5.3.1) with $p$ (5.3.8) and the initial conditions $(R_{\mathrm{in}},\dot{R}_{\mathrm{in}})$.

Again, determine $R_{\mathrm{in}}$ by the condition
\begin{equation} % 5.3.11
\dot{R}(0,t_{\mathrm{in}})=0
\end{equation}
An actual dynamics is this:
\begin{equation} % 5.3.12
R(t)=R(t,-\infty):=\lim_{t_{\mathrm{in}}\rightarrow-\infty}R(t,t_{\mathrm{in}})
\end{equation}

\section*{Acknowledgments}

I would like to thank Alex A. Lisyansky for support and Stefan V.
Mashkevich for helpful discussions.

\end{document}